# Atomic Structure of Graphene on SiO$_2$

Masa Ishigami[1,2], J.H. Chen[1,2], W.G. Cullen[1,2], M.S. Fuhrer[2,3] and E.D. Williams[1,2]

[1]*Materials Research Science and Engineering Center, University of Maryland, College Park, MD 20742 USA*

[2]*Department of Physics, University of Maryland, College Park, MD 20742, USA*

[3]*Center for Superconductivity Research, University of Maryland, College Park, MD 20742, USA*

## Abstract

We employ scanning probe microscopy to reveal atomic structures and nanoscale morphology of graphene-based electronic devices (i.e. a graphene sheet supported by an insulating silicon dioxide substrate) for the first time. Atomic resolution STM images reveal the presence of a strong spatially dependent perturbation, which breaks the hexagonal lattice symmetry of the graphitic lattice. Structural corrugations of the graphene sheet partially conform to the underlying silicon oxide substrate. These effects are obscured or modified on graphene devices processed with normal lithographic methods, as they are covered with a layer of photoresist residue. We enable our experiments by a novel cleaning process to produce atomically-clean graphene sheets.

Graphene[1,2], a single layer of graphite, is an unique material with exotic electronic properties[1-8]. A hexagonal two-dimensional network of carbon atoms composes graphene; it is exactly one atom in thickness and every carbon atom is a surface atom. Therefore, substrate-induced structural distortion[9], adsorbates[7], local charge disorder[10], atomic structure at the edges[4,11], and even atomic scale defects[12] could be very important for transport properties of graphene. Specifically, lowered carrier mobility[9] and suppression of weak localization[9] in graphene-based devices have been attributed to corrugation of the graphene. Consequently, understanding the atomic and nanoscale structures of graphene in the configuration in which it is measured is crucial to explaining the observed transport properties.

Experimentally, controlling the environment of graphene in a device configuration is difficult. Graphene on the common gate dielectric, $SiO_2$, is subject to the effects of trapped oxide charges[13], which are highly dependent on sample preparation. In addition, graphene devices are typically fabricated using electron beam lithography, exposing the graphene to photoresist that can leave behind contaminants which, like any chemical adsorbate, may modify electronic transport properties[10], may play a large role in reported graphene response to gas exposure[7], etc.. For instance, a freestanding graphene sheet has been reported to have intrinsic 3-D structure or ripples due to the instability of 2-D crystals[14,15]. However, the structures characterized had been exposed to photoresist, leaving the possibility that effects of chemical residues may have influenced the observed structure. Carefully controlling the experimental variables such as the influence of the substrate and the presence of impurities is necessary to interpret observed transport properties correctly.

In this letter, we report atomic structure and nanoscale morphology of monolayer graphene sheets and nanotubes in the most commonly used device configurations (i.e. on an insulating $SiO_2$ substrate with conducting back gate and fabricated electrical contacts). We find that acrylic lithography resists, commonly used in the device fabrication procedure, introduce unknown and uncontrollable perturbations, which must also apply to the majority of previously reported graphene-devices. The removal of the residue is necessary for uncovering intrinsic structural properties of the graphene sheet. Upon removing the resist residue, we are able to acquire atomic-resolution images of the graphene lattice, which shows both triangular and hexagonal lattice patterns in close proximity, indicating significant scattering of the electron waves. The atomic-resolution images also prove that our graphene devices are clean to atomic-scale, enabling controlled analysis of the structural properties. Finally, we measure the thickness of a graphene film in ultra high vacuum (UHV) and in ambient, and show that the large height measured in ambient is due to significant presence of atmospheric species under and/or on the graphene film.

We use scanning tunneling microscopy (STM) to achieve atomic-scale resolution, while we compare nanoscale morphologies of graphene and silicon dioxide substrate by non-contact atomic force microscopy (AFM). Unless otherwise noted, our microscopy studies were performed in UHV. In graphene devices, only the electronic contacts and graphene are conducting while the gate dielectric, which is insulating, composes the vast majority of the device substrate. Since STM requires conductive substrates, the STM tip must be positioned exactly above only the conductive areas, which extend laterally only several nanometers to microns for graphene devices. We use a commercial ultra high vacuum (UHV) system[16], which features a field emission scanning electron microscope (SEM) combined with AFM and STM for rapid, reproducible placement of scanned probe. Fig. 1a is an SEM image showing the scanned-probe tip approaching a representative carbon nanotube device to demonstrate our tip placement capability. In Fig. 1a, the nanotubes appear as thin curved white lines and the source/drain electrodes as the wider near-vertical lines, and a conductive AFM cantilever[17] is visible on the right. Coarse positioning of the cantilever within several microns of the nanotube is performed using SEM. We then utilize non-contact frequency-shift AFM[18] to locate the nanotube and to place the cantilever within several nanometers of the nanotube. Finally, the cantilever is employed as the STM tip; the tunneling current travels from the cantilever into the nanotube and along the nanotube into the electrodes. STM imaging is limited to the nanotube. As shown in Fig. 1b, this integrated technique is successful in resolving the atomic structure of nanotubes in the device configuration.

Fig. 2a is an AFM image of the graphene-based device, which we discuss in this paper. The wide white line, approximately 1 μm in width, is one electrode. The contacted graphitic material varies in the thickness but the large section appearing to the lower left is uniformly one monolayer thick, as will be shown later.

We find that a continuous film covers the surface of the graphene devices after the lift-off procedure, and it is not possible to obtain atomic resolution images via STM. A similar film was seen on the majority of nanotube segments in the nanotube devices, with only localized clean segments suitable for imaging. A control experiment using the same resist deposition[19] and acetone resist liftoff procedures on highly oriented pyrolytic graphite (HOPG) yields the same film, confirming its origin as residue from the resist. This indicates that the resist residue covers all graphene devices fabricated using similar photoresist process. Standard solvents such as Nano Remover PG[20] and glacial acetic acid do not perturb the residue. Known resist cleaning processes are inadequate for completely removing the resist residue.

We are able to remove the photoresist residue in argon/hydrogen atmosphere at 400 C[21]. Fig. 2c shows the AFM image of the same area shown in Fig. 2b, after the heat treatment. The graphene sheet now appears with finer, smoother corrugations. A representative large-area STM image of the cleaned graphene sheet is shown in Fig. 3. The atomic-scale pattern is visible in Fig. 3a, and can be imaged clearly at higher resolution as shown in Fig. 3b and 3d. The meandering of atomic rows seen in Fig. 3d is due to the curvature of the surface[22]. The observed lattice spacing is consistent with the graphene lattice, and the appearance of both triangular and hexagonal lattice in the image indicates the presence of strong spatially dependent perturbations which interact with graphene electronic states[23,24]. Such perturbations may be due to the observed film curvature and/or the charge traps on the $SiO_2$ surface. Significantly, STM images at any position on the device always reveal the graphitic lattice. Therefore, surface impurities have been removed completely from the graphene surface, and the corrugation seen in Fig. 2c is representative of the clean graphene sheet on $SiO_2$.

The material thickness is one of the key structural factors in determining the properties of graphene-based devices[3]. Fig. 4a shows an AFM image of the boundary between the same graphene sheet and $SiO_2$ substrate. A histogram acquired across the boundary shown in Fig. 4b shows that the film thickness is 4.2 Å, comparable to the layer-to-layer spacing in bulk graphite of 3.4 Å. Therefore, the imaged graphene device area is a monolayer. Similar analysis performed in air for the same area, before our experiments in UHV, shows the thickness to be 9 Å, consistent with a previous measurement of a monolayer material in air[1,8]. The discrepancy between the air/vacuum measurements of 4.6 Å in thickness indicates a significant presence of ambient species (nitrogen, oxygen, argon, or water) between $SiO_2$ and the graphene sheet and/or on the graphene sheet.

We now turn our attention to the 3-D morphology of the graphene sheet, important for the transport properties[9]. Fig. 4c shows histograms of the heights over graphene and $SiO_2$. The graphene sheet is approximately 60% smoother than the oxide surface; the standard deviations of the measured height variations are 1.9 Å and 3.1 Å for the graphene and oxide surface. The height-height correlation function, $g(x) = \langle (z(x_0 + x) - z(x_0))^2 \rangle$, is a useful measure for characterizing the surface morphology[25-27]. Fig. 4d shows the height-height correlation function[28] for the graphene and $SiO_2$ surface. Both correlation functions rapidly increase as $g \sim x^{2H}$ at short distances, as expected[26]: $2H = 1.11 \pm 0.013$ for graphene and $2H = 1.17 \pm 0.014$ for $SiO_2$. A value of the exponent $2H \sim 1$ indicates a domain structure with short-range correlations among neighboring domains[25] and is not surprising for $SiO_2$. A value of $2H = 2$ is expected[29] for a thermally-excited flexible membrane under the influence of an interaction (e.g. van der Waals) with the substrate. Consequently, the observed 2H value demonstrates that the observed graphene morphology is not representative of the intrinsic structure. A rollover at the correlation length and saturation at mean square roughness at large distances follow the short-distance behavior. As seen in the figure inset, interpolating the intersection of the power-law and saturated regimes yields values of the correlation length[26], which are $\xi = 32 \pm 1$ nm for graphene and $\xi = 23 \pm 0.6$ nm for $SiO_2$. The similar exponents and slightly larger correlation length of the graphene sheet is consistent with the graphene morphology being determined by the underlying $SiO_2$ substrate. The larger correlation length and smaller roughness of the graphene surface would arise naturally due to an energy cost for graphene to closely follow sharp orientation changes on the substrate. Freestanding graphene has been reported to have larger static nanoscale corrugations[14,15] attributed to intrinsic structural instability of 2D materials. However, the free-standing graphene was treated with a resist process[14,15], and the resulting resist residue could certainly prevent the graphene sheet from reaching its equilibrium structural corrugation.

The observed corrugations in our study indicate a maximum local strain of approximately 1 %. Using the Young's modulus of 1 TPa[30] and graphene thickness of 3.4 Å, the corresponding stored energy density due to the induced deformation is ~1 meV/Å$^2$. We estimate the graphene-SiO$_2$ interaction energy to be >6 meV/Å$^2$ based on the interlayer van der Waals interaction in graphite[31] of 20 meV/Å$^2$ at the distance of 3.4 Å. The estimated interaction energy between the graphene sheet and SiO$_2$ substrate is thus sufficient to overcome the energy cost of the corrugations needed for graphene to follow the SiO$_2$ morphology.

Corrugations comparable to those observed here have been postulated to be responsible for the lack of low-field magnetoresistance observed in graphene on SiO$_2$ via suppression of weak localization due to the introduction of an effective random magnetic field[9]. Indeed, "flatter" graphene films, prepared on SiC with the film coherence length of 90 nm, show weak localization[2]. The corrugations in graphene on SiO$_2$ were later attributed to intrinsic corrugations in the graphene itself[15]. However, our findings indicate that the graphene corrugations that are relevant for interpreting many reports of device performance (e.g. for graphene on SiO2) are due to partial conformation of the graphene to the SiO$_2$, not to the intrinsic corrugation of graphene.

We have resolved atomic structures of oxide-supported graphene-based electronic devices using a novel combined SEM/AFM/STM technique. We obtain real-space images of the single-layer graphene atomic lattice for the first time, and characterize the thicknesses and nanoscale corrugation of a clean graphene sheet devoid of any impurities. Our observation shows that the graphene primarily follows the underlying morphology of SiO$_2$ and thus does not have intrinsic, independent corrugations on SiO$_2$. The graphene sheets *do* have finite intrinsic stiffness, which prevents the sheets from conforming completely to the substrate. In addition, we demonstrate that resist residues are ubiquitous on lithographically-fabricated graphene devices, and their presence should be considered in interpreting transport and structural measurements of earlier studies. Our quantitative measure of the extrinsic corrugations of graphene on SiO$_2$ can be used as input to theoretical models of strain-induced disorder in graphene and its effect on transport properties. Furthermore, our observation that graphene can conform to substrate morphology suggests new experimental directions: the use of controlled substrate morphologies (e.g. a patterned $SiO_2$ substrate, or alternative dielectric materials) may be a useful approach to investigate how the corrugation-induced strain impacts the transport properties of graphene. Finally, our technique (the novel integrated microscopy allied with the resist cleaning process) can be applied to resolve atomic structures of nanoelectronic devices in general; the technique finally enables studies of the impact of atomic scale defects and adsorbates on nanoscale transport properties.

**Acknowledgments**

This work has been supported by the Director of Central Intelligence Postdoctoral Fellowship program, the Laboratory for Physical Sciences, the U.S. Office of Naval Research grant no. N000140610882, and NSF grant no. CCF-06-34321. Use of the UHV microscopy and nanofabrication facilities is supported by the UMD-NSF-MRSEC under grant DMR 05-20471, and the initial purchase of the SEM/STM instrument was supported by the NSF-IMR program.

19. For this control experiment, we tested both polymethyl methacrylate (PMMA) and copolymer PMMA/8.5% methacrylic acid (MMA-MAA) resists in contact with the substrate. Both resists show similar resist residues upon lift-off.
20. N-methyl-2-pyrrolidone (NMP)-based remover sold by Microchem.
21. The flow rates were 1700 ml/min for argon and 1900 ml/min for hydrogen. Our anneal time was 1 hour. Ar/H2 flow must be maintained throughout the heating and annealing. Otherwise, the residue film is not removed. .
22. Venema, L. C., Menueir, V., Lambin, P. & Dekker, C. Atomic structure of carbon nanotubes from scanning tunneling microscopy. *Phys. Rev. B* **61**, 2991 (2000).
23. Tomanek, D. et al. Theory and observation of highly asymmetric atomic structure in scanning-tunneling-microscopy images of graphite. *Physical Review B* **35**, 7790 (1987).
24. Kane, C. L. & Mele, E. J. Broken symmetries in scanning tunneling images of carbon nanotubes. *Physical Review B* **59**, R12759 (1999).
25. Goldberg, J. L., Wang, X.-S., Bartelt, N. C. & Williams, E. D. Surface height correlation functions of vicinal Si(111) surfaces using scanning tunneling microscopy. *Surface Science Letters* **249**, L285-92 (1991).
26. Krim, J., Heyvaert, I., Haesendonck, C. V. & Bruynseraede, Y. Scanning Tunneling Microscopy Observation of Self-Affine Fractal Roughness in Ion-Bombarded Film Surfaces. *Physical Review Letters* **70**, 57 (1993).
27. Palasantzas, G., Tsamoras, D. & De Hosson, J. T. M. Roughening aspects of room temperature vapor deposited oligomer thin films onto Si substrate. *Surface Science* **507-10**, 357-361 (2002).
28. For each correlation function, we average seven different horizontal line profiles acquired on each respective surface. Constant linear offsets at large distances from 50 to 150 nm have been subtracted to account for imperfect plane tilt correction performed on the image. .
29. Aranda-Espinoza, H. & Lavallee, D. Structure factor of flexible membranes. *Europhysics Letters* **43**, 355-359 (1998).
30. Krishnan, A., Dujardin, E., Ebbesen, T. W., Yianilos, P. N. & Treacy, M. M. J. Young's modulus of single-walled nanotubes. *Physical Review B (Condensed Matter)* **58**, 14013-14019 (1998).
31. Chopra, N. G. et al. Fully collapsed carbon nanotubes. *Nature* **377**, 135-138 (1995).
32. Kim, W. K. et al. Synthesis of ultralong and high percentage of semiconducting single-walled carbon nanotubes. *Nano Letters* **2**, 703 (2002).
33. In the first EBL step, alignment markers are defined using EBL near the monolayer materials. Electronic contacts are then defined in the second EBL step using the alignment markers. The contacts are gold with chromium sticking layer. Typically, PMMA/MMA-MAA bilayer resist is used for EBL.

**Figures**

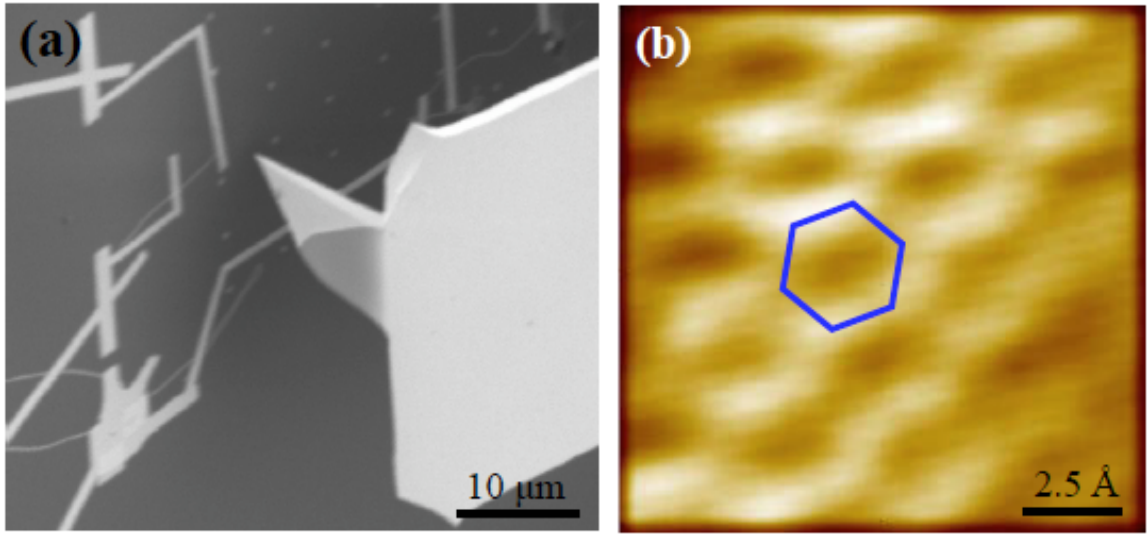

Figure 1(a) Scanning electron micrograph of a carbon nanotube device, showing our experimental setup. The triangular shape to the right of the image is the tip of the scanning probe. The nanotubes were grown using chemical vapor deposition following reference[32] and the electrodes were patterned using a standard two-step electron-beam lithography process[33] The device substrate is 500 nm thick thermal $SiO_2$ grown on a heavily-doped silicon wafer. Wide, near vertical lines on the left are electrical contacts. Thin white lines are the nanotubes lying on the surface of $SiO_2$. (b) An STM image of a nanotube in the device configuration, showing atomic structure. $V_{source}=V_{drain}=1.4$ v, $V_{gate}=0$ v, and $I_{tunnel}=18$ pA.

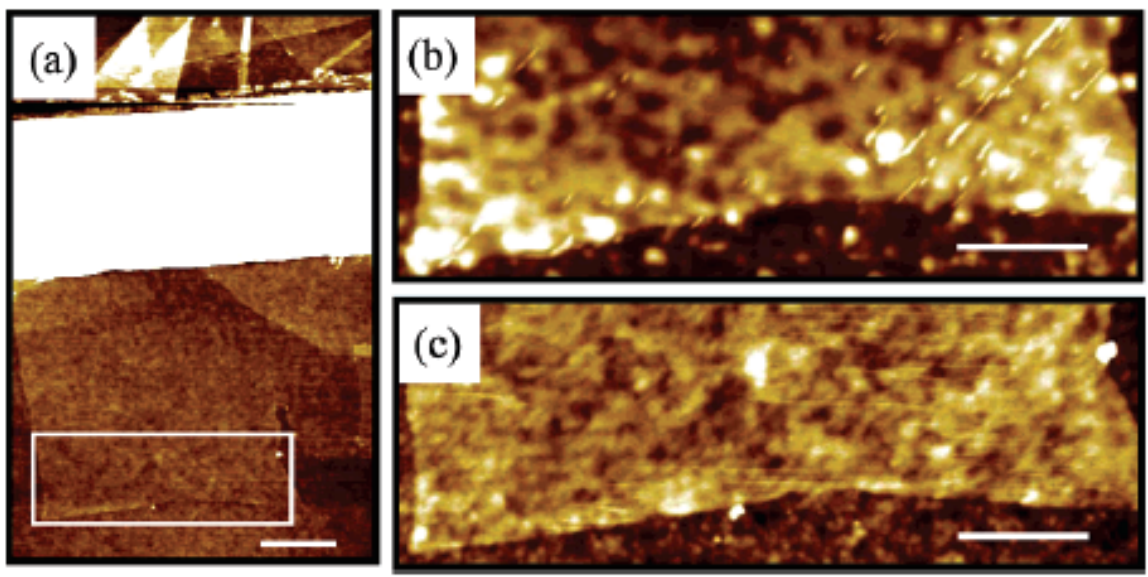

Figure 2 (a) AFM topography of graphene deposited on $SiO_2$. Thin graphite flakes are generated using the mechanical exfoliation technique[1] on thermally grown $SiO_2$ with the thickness of 300 nm. Monolayer graphite flakes (graphene) are located using optical and atomic force microscopy[8]. The e-beam lithography defined electrode[33], approximately 80 nm in height and 1.5 μm in width, is the white area nearly horizontal to the image. The black square indicates the region shown in Figs. 1b and c. The scale bar is 500 nm. (b) Graphene sheet prior to the cleaning procedure described in text. The scale bar is 300 nm. (c) Graphene sheet after the cleaning procedure. The standard deviation of the height variation in a square of side 600 nm is approximately 3 Å after the treatment compared to 8 Å before the treatment. The scale bar is 300 nm. Images (a)-(c) were acquired using intermittent-contact mode AFM in air.

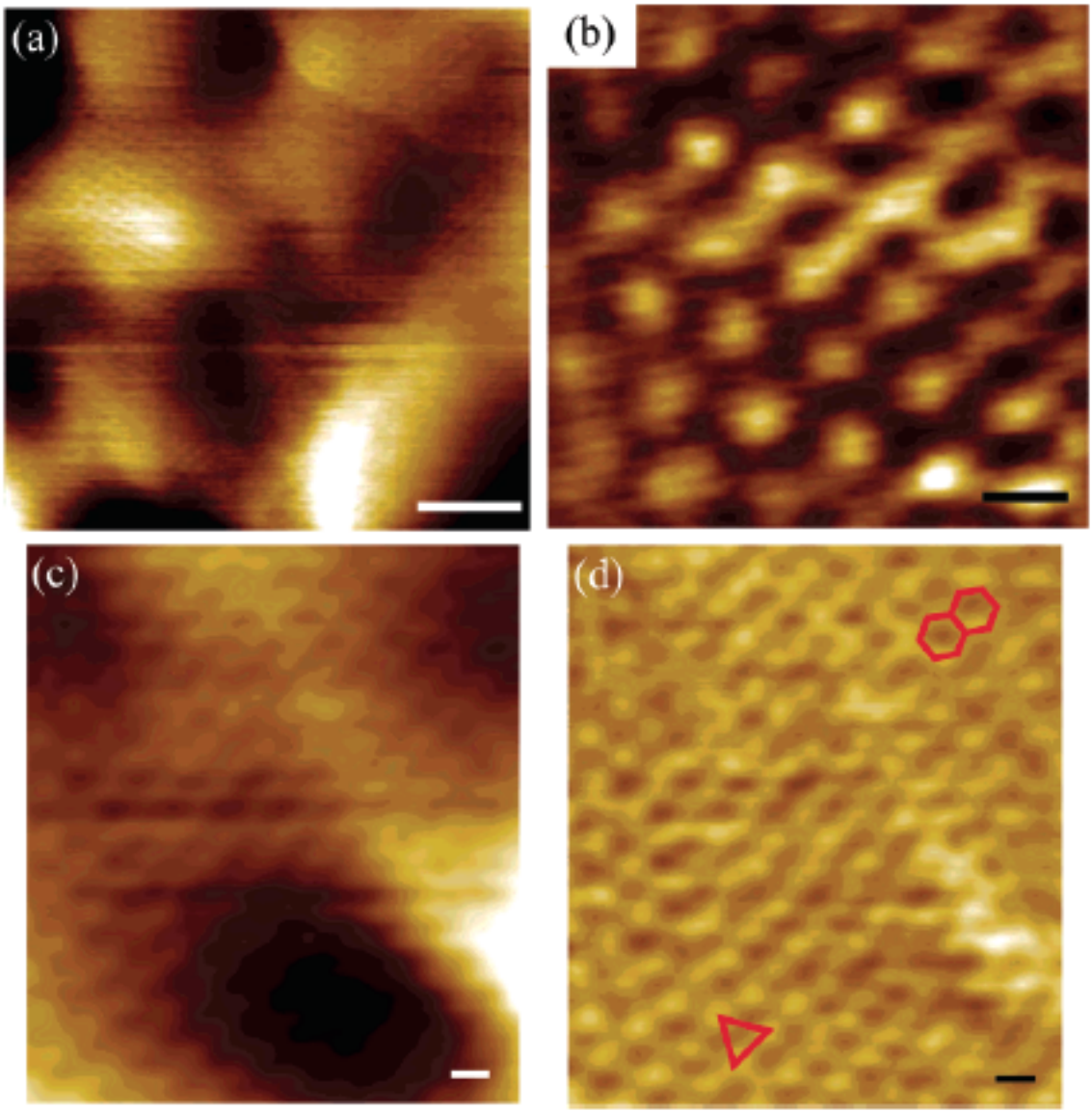

Figure 3 (a) A typical large-area STM image of the graphene sheet shown in Fig. 2a. Peak-to-peak height variation of the image is approximately 2.5 nm. $V_{sample}$ = 1.1 V and $I_{tunnel}$ = 0.3 nA. The scale bar is 2 nm. (b) Atomically-resolved image of a graphene sheet. $V_{sample}$ = 1.0 V and $I_{tunnel}$ = 24 pA. The scale bar is 2.5 Å. (c) STM image of another area. The scale bar is 2.5 Å. $V_{sample}$ = 1.2v and $I_{tunnel}$ = 0.33 nA. (d) A high-pass filtered image of the large area scan shown in (c). Both triangular and hexagonal patterns are observed. The orientations of the red triangle and hexagons are same. The scale bar is 2.5 Å.

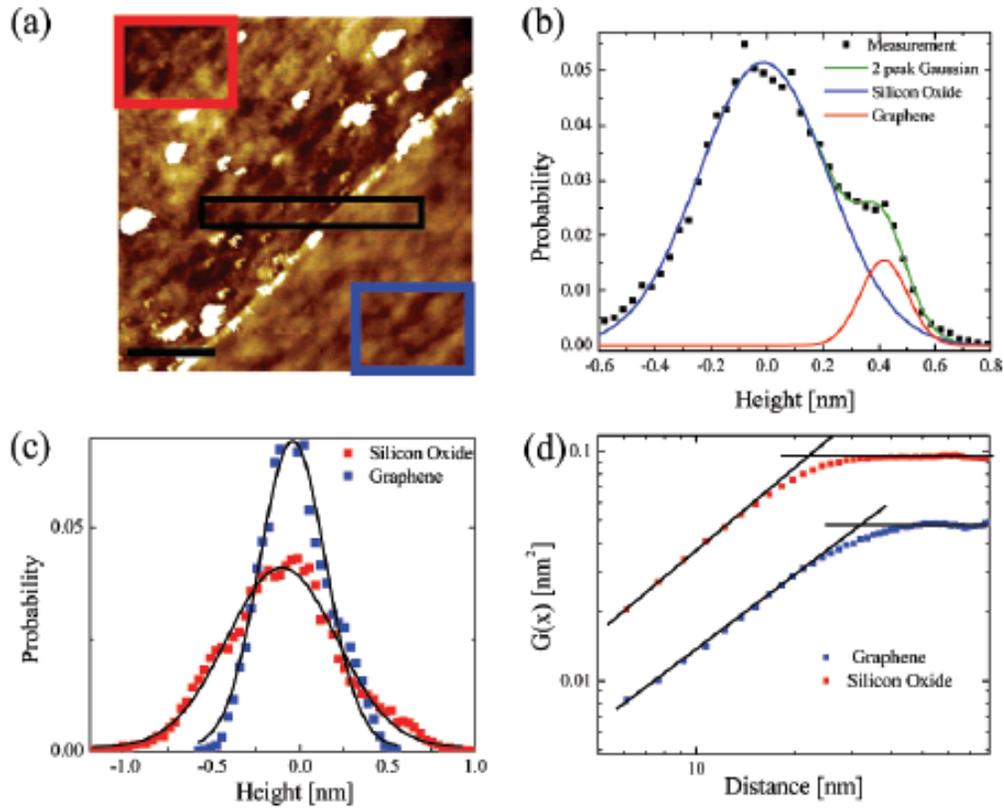

Figure 4 (a) Non-contact mode AFM image, acquired in UHV, of a boundary between the graphene sheet and SiO$_2$ substrate. The graphene sheet occupies the lower right area of the image. The scale bar is 200 nm. The black rectangle indicates the area for the histogram shown in Fig. 4b, and red and blue rectangles indicate the area where the histograms shown in Fig. 4c has been acquired. (b) Height histogram acquired across the graphene-substrate boundary (black rectangle in Fig. 4a). The data are fit by two Gaussian distributions (solid red and blue lines; green line is sum), with means separated by 4.2 Å. (c) Height histograms acquired on graphene and SiO$_2$ (red and blue squares respectively in Fig. 4a). The histograms are well-described by Gaussian distributions (black lines) with standard deviations of 1.9 Å and 3.1Å for graphene and SiO$_2$, respectively. (d) The height-height correlation function (see text) of the graphene sheet and SiO$_2$ surface. The lines are fits to the large and small length behaviors (power-law and constant, respectively), and the point of intersection indicates the correlation length. This analysis is performed by selecting data from Fig. 4a, showing both graphene and SiO$_2$ surfaces. Therefore, the tip morphology is the same for both curves and the tip-related artifact effect does not contribute to the analysis.